\documentclass[preprint,superscriptaddress,nofootinbib]{revtex4}
  
  \usepackage{amssymb,amsmath,graphicx,subfigure,color}
  \usepackage[colorlinks]{hyperref}
  \allowdisplaybreaks[4]
  \begin{document}

  \title{Feasibility of the experimental study of
         $D_{s}^{\ast}$ ${\to}$ ${\phi}{\pi}$ decay}
  \author{Yueling Yang}
  \affiliation{Institute of Particle and Nuclear Physics,
              Henan Normal University, Xinxiang 453007, China}
  \author{Kang Li}
  \affiliation{Institute of Particle and Nuclear Physics,
              Henan Normal University, Xinxiang 453007, China}
  \author{Zhenglin Li}
  \affiliation{Institute of Particle and Nuclear Physics,
              Henan Normal University, Xinxiang 453007, China}
  \author{Jinshu Huang}
  \affiliation{School of Physics and Electronic Engineering,
              Nanyang Normal University, Nanyang 473061, China}
  \author{Junfeng Sun}
  \affiliation{Institute of Particle and Nuclear Physics,
              Henan Normal University, Xinxiang 453007, China}

  \begin{abstract}
  The current knowledge on the $D_{s}^{\ast}$ meson are
  very limited. Besides the dominant electromagnetic decays,
  the $D_{s}^{\ast}$ weak decays are legal and offer the valuable
  opportunities to explore the wanted $D_{s}^{\ast}$ meson.
  In this paper, the $D_{s}^{\ast}$ ${\to}$ ${\phi}{\pi}$ decay
  was studied with the factorization approach.
  It is found that the branching ratio
  ${\cal B}(D_{s}^{\ast}{\to}{\phi}{\pi})$ ${\sim}$
  ${\cal O}(10^{-7})$, which corresponds to several
  thousands of events at the $e^{+}e^{-}$ collider
  experiments including STCF, SuperKEKB, CEPC and
  FCC-ee, and several millions of events at the
  hadron collider experiments, such as LHCb@HL-LHC.
  It is feasible to experimentally study the
  $D_{s}^{\ast}$ ${\to}$ ${\phi}{\pi}$ weak decay
  in the future, even considering the
  identification efficiency.

  \href{https://doi.org/10.1140/epjc/s10052-022-10524-5}{Eur. Phys. J. C 82, 555 (2022)}

  \end{abstract}

  \maketitle

  \section{Introduction}
  \label{sec01}
  The first evidence for the charmed strange mesons
  $D_{s}^{\ast}$ was observed in the exclusive reaction
  $e^{+}e^{-}$ ${\to}$ $F\bar{F}^{\ast}$ by the
  DASP collaboration in the year of 1977 \cite{Phys.Lett.B.70.132},
  where the symbols of $F$ and $F^{\ast}$ were formerly used to
  denote the $D_{s}$ and $D_{s}^{\ast}$ particles, respectively.
  According to the $SU(4)$ quark model assignments, the vector
  mesons $D_{s}^{\ast}$ are assumed to have the same
  quark compositions as their twin pseudoscalar partners
  $D_{s}$.
  Both the $D_{s}^{{\ast}+}$ and $D_{s}^{+}$ mesons are consisting
  of a quark-antiquark pair $c\bar{s}$, and have the same additive quantum
  numbers of Charm, Strangeness and Charge, {\em i.e.}, $C$ $=$
  $S$ $=$ $Q$ $=$ $+1$.
  The different spin configurations of interquark potential
  make the mass of the ground spin-triplet $1^{3}S_{1}$ state
  for the $D_{s}^{\ast}$ mesons to be above that of the ground
  spin-singlet $1^{1}S_{0}$ state for the $D_{s}$ mesons
  \cite{pdg2020}.

  Compared with the pseudoscalar meson $D_{s}$, the
  experimental information on the properties of the vector meson
  $D_{s}^{\ast}$ is still very limited by now \cite{pdg2020}.
  Although there were many measurements of the mass of the
  $D_{s}^{\ast}$ meson (such as in Refs. \cite{Phys.Lett.B.70.132,
  Phys.Lett.B.80.412,Phys.Lett.B.146.111,PhysRevLett.53.2465,
  PhysRevLett.58.2171,Phys.Lett.B.207.349,PhysRevD.50.1884,
  PhysRevLett.75.3232}),
  only one measurement was solemnly quoted
  by the Particle Data Group (PDG) until now \cite{pdg2020}.
  The measurement was carried out by the Mark III collaboration
  in 1987 \cite{PhysRevLett.58.2171}, thirty-five years ago.
  And the errors of the measurement of mass, $m_{D_{s}^{\ast}}$ $=$
  $2109.3{\pm}2.1{\pm}3.1$ MeV \cite{PhysRevLett.58.2171},
  are significantly larger than those of current values
  of the $D_{s}$ meson, $m_{D_{s}}$ $=$ $1968.35{\pm}0.07$ MeV \cite{pdg2020}.
  For the full width of the $D_{s}^{\ast}$ meson, only the upper
  limit was given by different experimental groups \cite{pdg2020}
  and the latest and minimal upper limit on the decay width of the
  $D_{s}^{\ast}$ meson was given by the CLEO Collaboration
  in 1995 \cite{PhysRevLett.75.3232}, twenty-seven years ago.
  The natural spin-parity of the $D_{s}^{\ast}$ meson was analyzed
  to be most likely $J^{P}$ $=$ $1^{-}$ \cite{PhysRevLett.75.3232},
  but has not been unambiguously determined experimentally \cite{pdg2020}.

  The experimental data on the $D_{s}^{\ast}$ mesons are
  accumulating increasingly.
  The quantitative study on the $D_{s}^{\ast}$ mesons is
  coming.
  Inspired by the potential prospects of high-luminosity-frontier
  flavor experiments, more and more data of the $D_{s}^{\ast}$
  mesons will be available, so more accurate information and
  more detailed knowledge of the properties of the $D_{s}^{\ast}$
  mesons will be accessible.
  In the $e^{+}e^{-}$ colliders, it is promisingly
  expected that there will be a total of about $5{\times}10^{10}$
  $c\bar{c}$ pairs at the SuperKEKB \cite{PTEP.2019.123C01},
  about $10^{11}$ $c\bar{c}$ pairs from $10^{12}$ $Z^{0}$ boson decays
  at the Circular Electron Positron Collider (CEPC) \cite{cepc},
  about $6{\times}10^{11}$ $c\bar{c}$ pairs from $5{\times}10^{12}$
  $Z^{0}$ boson decays at the Future Circular Collider (FCC-ee) \cite{fcc},
  where the branching fraction for the $Z^{0}$ boson decay into
  the $c\bar{c}$ pair is ${\cal B}(Z^{0}{\to}c\bar{c})$ $=$
  $(12.03{\pm}0.21)\%$ \cite{pdg2020}.
  Considering the fraction of the charmed quark fragmenting into the
  $D_{s}^{\ast}$ meson $f(c{\to}D_{s}^{\ast})$ ${\simeq}$
  $5.5\%$ \cite{epjc.76.397}, these high statistical $c\bar{c}$ pairs
  correspond to some $6{\times}10^{9}$, $10^{10}$ and
  $6{\times}10^{10}$ $D_{s}^{\ast}$ mesons at the SuperKEKB,
  CEPC and FCC-ee, respectively.
  In addition, about $10^{10}$ $D_{s}^{\ast}$ mesons are expected
  above the ${\psi}(4040)$ threshold (see Fig. 6 of
  Ref. \cite{epjc.81.1110})
  at both the super ${\tau}$-charm factory (STCF)
  in China \cite{STCF} and the super charm-tau factory (SCTF) in
  Novosibirsk, Russia \cite{SCTF}, based on an integrated
  luminosity of $10\,{ab}^{-1}$ \cite{STCF}.
  In the high-energy hadron colliders, about $4{\times}10^{13}$
  $D_{s}^{\ast}$ mesons \cite{epjc.81.1110} are expected to be
  obtainable with a data sample of target luminosity $300\,fb^{-1}$
  at the LHCb@HL-LHC experiments \cite{epjst.228.1109},
  and more $D_{s}^{\ast}$ mesons will be accumulated at ALICE
  and ATLAS \cite{epjc.81.1110}.
  The huge amount of experimental data provide a tremendous
  foundation and valuable opportunities for studying and
  understanding the properties of $D_{s}^{\ast}$ meson.
  A brilliant portrait of the characteristics of $D_{s}^{\ast}$
  mesons is going to be unfolded smoothly and completely.

  The fit mass of $D_{s}^{\ast}$ meson is $m_{D_{s}^{\ast}}$
  $=$ $2112.2{\pm}0.4$ MeV \cite{pdg2020}, just below the mass
  threshold of the $D\overline{K}$ pair and above the mass
  threshold of the $D_{s}{\pi}$ pair and , {\em i.e.}, the mass
  relations $m_{D_{u,d}}$ $+$ $m_{K}$ $>$ $m_{D_{s}^{\ast}}$ $>$
  $m_{D_{s}}$ $+$ $m_{\pi}$.
  Thus the hadronic decays $D_{s}^{\ast}$ ${\to}$ $D\overline{K}$
  are strictly forbidden by the law of conservation of energy.
  The hadronic decay $D_{s}^{\ast}$ ${\to}$ $D_{s}{\pi}$ is
  permissible kinematically, but violates the the isospin
  conservation in the strong interactions\footnotemark[1].
  \footnotetext[1]{Within the chiral perturbative theory,
  it is usually taken for granted that the $D_{s}^{\ast}$ ${\to}$
  $D_{s}{\pi}$ decay can also decay through the strong interactions
  via the ${\pi}^{0}$-${\eta}$ mixing by assuming a small isocalar
  ${\eta}$ meson component in the physical ${\pi}^{0}$ meson,
  because the ${\eta}$ meson can couple to the strange quark
  in the charmed strange mesons \cite{PhysRevD.49.6228,
  Nucl.Phys.B.529.62,Nucl.Phys.A.710.99,PhysRevD.101.054019}.}
  The absences of decay modes induced by the strong interactions
  make the $D_{s}^{\ast}$ meson to be very narrow.
  The natural width of the the $D_{s}^{\ast}$ meson is significantly
  less than the best experimental resolution.
  Here, it should be noted that
  the $D_{s}^{\ast}$ ${\to}$ $D_{s}{\pi}$ decay
  is suppressed not only by the phenomenological Okubo-Zweig-Iizuka
  (OZI) rule \cite{ozi-o,ozi-z,ozi-i} but also by the extremely
  limited phase spaces, due to $m_{D_{s}^{\ast}}$ $-$ $m_{D_{s}}$
  $-$ $m_{\pi}$ $<$ $6$ MeV.
  Thus the electromagnetic decay $D_{s}^{\ast}$ ${\to}$
  $D_{s}{\gamma}$ is dominant, with the branching ratio
  ${\cal B}(D_{s}^{\ast}{\to}D_{s}{\gamma})$ $=$ $(93.5{\pm}0.7)\%$
  exceeding that of hadronic decay
  ${\cal B}(D_{s}^{\ast}{\to}D_{s}{\pi})$ $=$ $(5.8{\pm}0.7)\%$
  \cite{pdg2020}.
  In addition, for the $D_{s}^{\ast}$ ${\to}$ $D_{s}{\pi}^{0}$,
  $D_{s}{\gamma}$ decays\footnotemark[2]
  \footnotetext[2]{The neutral pion decay predominantly through
  ${\pi}^{0}$ ${\to}$ ${\gamma}{\gamma}$  with a branching
  ratio of $98.8\%$ \cite{pdg2020}.}, the final photons are
  seriously polluted by those from bremsstrahlung radiation,
  which will significantly affect the
  identification efficiency of the accident photon.
  Besides, the $D_{s}^{\ast}$ meson can also decay via the weak
  interactions, although with a very small probability.
  The weak decays of the $D_{s}^{\ast}$ meson provide another
  platform and opportunities to explore and understand the
  properties of the $D_{s}^{\ast}$ mesons.
  In this paper, we will evaluate the feasibility of experimentally
  investigating the $D_{s}^{\ast}$ meson through the weak decay
  $D_{s}^{\ast}$ ${\to}$ ${\phi}{\pi}$.

  Theoretically, the charm-flavor-changing decay $D_{s}^{{\ast}+}$
  ${\to}$ ${\phi}{\pi}^{+}$ is actually induced by the quark transition
  $c$ ${\to}$ $s$ $+$ $W^{{\ast}+}$ at the tree level
  in the standard model (SM) of elementary particles.
  Here, it is assumed that the vector ${\phi}$ meson consists of
  the pure $s\bar{s}$ quark pair with neither possible $u\bar{u}$ nor
  $d\bar{d}$ components, {\em i.e.}, that the mixing between
  the ${\phi}$-${\omega}$ system is ideal.
  Clearly, this decay mode is the Cabibbo-favored one and its
  amplitudes are proportional to the Cabibbo-Kobayashi-Maskawa
  (CKM) matrix \cite{PhysRevLett.10.531,PTP.49.652} element
  ${\vert}V_{cs}{\vert}$ ${\sim}$ ${\cal O}(1)$.
  This decay would have a relatively large branching ratio
  among the $D_{s}^{\ast}$ meson weak decays, and hence
  should have a high priority to be studied.
  In addition, the charm quark is somewhat massive and can be
  regarded as one bridge between the perturbative and
  nonperturbative regimes.
  The charm quark decays offer a laboratory to test various
  phenomenological models and study the behaviors of the
  strong interactions near the scale of ${\cal O}(m_{c})$.

  Experimentally, the curved tracks of the charged pion and kaon
  plunged into magnetic field will be unambiguously detectable by
  the highly sensitive detectors.
  So, the final states are easily identified for the
  $D_{s}^{\ast}$ ${\to}$ ${\phi}{\pi}$ decays, where ${\phi}$ and
  ${\pi}$ mesons with a definite momentum are back-to-back in the
  center-of-mass frame of the $D_{s}^{\ast}$ meson, and the
  ${\phi}$ meson can be well reconstructed from the kaon pairs.
  It is expected to have a higher signal-to-background ratio
  and a better identification efficiency, and have a big
  competitive advantage over both the pure leptonic decays
  $D_{s}^{\ast}$ ${\to}$ ${\ell}\bar{\nu}$ and semileptonic
  decays $D_{s}^{\ast}$ ${\to}$ ${\phi}{\ell}\bar{\nu}$
  which suffer from the additional complications caused
  by the final neutrinos.

  In this paper, we will study the $D_{s}^{\ast}$ ${\to}$
  ${\phi}{\pi}$ decay within SM by using the
  phenomenological factorization
  approach \cite{zpc.34.103}, and estimate the branching
  ratio in order to provide a ready reference for future
  experimental analysis.
  This paper is organized as follows.
  The amplitudes for the $D_{s}^{\ast}$ decay in question
  using the factorization approximation is given in
  Sec. \ref{sec:hamiltonian}.
  Branching ratio and event numbers of the $D_{s}^{\ast}$
  ${\to}$ ${\phi}{\pi}$ decay are listed in Sec. \ref{sec:result}.
  Section \ref{sec:summary} devotes to a summary.

   \section{The theoretical framework}
   \label{sec:hamiltonian}
   At the quark level, the effective Hamiltonian
   responsible for the nonleptonic  decay $D_{s}^{\ast}$
   ${\to}$ ${\phi}{\pi}$ can be written as
   \cite{RevModPhys.68.1125},
   \begin{equation}
  {\cal H}_{\rm eff}\, =\,
   \frac{G_{F}}{\sqrt{2}}\, V_{cs}^{\ast}\,V_{ud}\, \big\{
    C_{1}\,O_{1}+C_{2}\,O_{1} \big\} + {\rm h.c.}
   \label{hamilton},
   \end{equation}
   where the Fermi constant $G_{F}$ is the weak interaction
   coupling coefficient, $G_{F}$ ${\approx}$
   $1.166{\times}10^{-5}$ ${\rm GeV}^{-2}$ \cite{pdg2020}.
   $V_{cs}^{\ast}\,V_{ud}$ is the product of CKM
   matrix elements, which has been determined
   precisely by experiments,
   ${\vert}V_{ud}{\vert}$ $=$ $0.97370(14)$ and
   ${\vert}V_{cs}{\vert}$ $=$ $0.987(11)$ \cite{pdg2020}.
  The Wilson coefficients $\vec{C}$ $=$ $\{C_{1},C_{2}\}$
  can be obtained with the renormalization group equation,
  \begin{equation}
  \vec{C}({\mu}_{c}) \, =\,
  U_{4}({\mu}_{c},m_{b})\, M(m_{b})\,
  U_{5}(m_{b},m_{W})\, \vec{C}(m_{W})
  \label{ci},
  \end{equation}
  where ${\mu}_{c}$ ${\sim}$ ${\cal O}(m_{c})$ is the scale
  for the charm quark decays.
  $m_{b}$ and $m_{W}$ are the mass of the bottom quark and
  the charged $W$ gauge boson, respectively.
  $U_{f}({\mu}_{f},{\mu}_{i})$ and $M(m_{b})$ are the
  evolution matrix and threshold matching matrix, respectively.
  The expressions of $\vec{C}(m_{W})$, $U_{f}({\mu}_{f},{\mu}_{i})$
  and $M(m_{b})$ 
  can be found in Ref. \cite{RevModPhys.68.1125}.
  The effective operators are defined as follows.
   \begin{eqnarray}
   O_{1} &=&
   \big[ \bar{s}_{\alpha}\,{\gamma}^{\mu}\,(1-{\gamma}_{5})\,c_{\alpha} \big]\,
   \big[ \bar{u}_{\beta}\,{\gamma}_{\mu}\,(1-{\gamma}_{5})\,d_{\beta} \big]
   \label{operator-o1},  \\
   O_{2} &=&
   \big[ \bar{s}_{\alpha}\,{\gamma}^{\mu}\,(1-{\gamma}_{5})\,c_{\beta} \big]\,
   \big[ \bar{u}_{\beta}\,{\gamma}_{\mu}\,(1-{\gamma}_{5})\,d_{\alpha} \big]
   \label{operator-o2},
   \end{eqnarray}
   where ${\alpha}$ and ${\beta}$ are the color indices.
   Because the $D_{s}^{\ast}$ ${\to}$ ${\phi}{\pi}$ decay
   is an external $W$ emission process, there are only
   two current-current operator $O_{1,2}$ and without
   the penguin operators, and the contributions from
   new physics beyond SM to this decay are negligible.

   The initial and final states are hadrons, while the
   operators are the specific combinations of four quarks.
   The influence of the long-distance strong interactions
   on the transitions between quarks and hadrons makes the
   predictions of nonleptonic decays notoriously difficult.
   To obtain the decay amplitudes for the $D_{s}^{\ast}$
   ${\to}$ ${\phi}{\pi}$ decay, the remaining work is to
   evaluate the hadronic matrix elements (HMEs)
   ${\langle}{\phi}{\pi}{\vert}O_{i}{\vert}D_{s}^{\ast}{\rangle}$.

   Phenomenologically, one of the most frequently used methods
   to deal with HME is the naive factorization (NF) approach
   \cite{zpc.34.103}.
   The NF approach is based on the color transparency
   hypothesis \cite{npbps.11.325} that a nearly collinear
   and relativistic light quark-antiquark pair originating from
   the heavy quark decays might be approximated as a color
   singlet before its hadronization and complete separation
   from the interaction points.
   According to the color transparency hypothesis, it is possible
   to replace the product of the quark currents in the effective
   Hamiltonian of Eq.(\ref{hamilton}) by product of the
   corresponding hadron currents, and express the
   color singlet quark currents in terms of the
   participating hadron fields \cite{Stech.1985}.
   The outgoing light hadrons of two-body decays
   are back-to-back and energetic in the heavy quark limit,
   and fly away far from each other before the interference
   with the soft gluons.
   It may be a good approximation to neglect the final state
   interactions for the moment.
   In addition, the asymptotic freedom property of the strong
   interactions implies that the creation of quark pairs of
   high energy from the vacuum by hard virtual gluon is highly
   suppressed  \cite{npb.133.315}, {\em i.e.}, it is believed
   that the $W$-annihilation amplitudes for the nonleptonic
   heavy-flavored hadron decays might be much smaller than
   the $W$-emission amplitudes.

   Under the assumption of factorization, the decay
   amplitudes are written as,
   \begin{eqnarray}& &
  {\cal A}(D_{s}^{\ast}{\to}{\phi}{\pi})\, =\,
  {\langle}{\phi}\,{\pi}{\vert}{\cal H}_{\rm eff}
  {\vert}D_{s}^{\ast}{\rangle}
   \nonumber \\ &=&
   \frac{G_{F}}{\sqrt{2}}\, V_{cs}^{\ast}\,V_{ud}\,
   a_{1}\, {\langle}{\phi}\,{\pi}{\vert}
   (\bar{s}\,c)_{H}\, (\bar{u}\,d)_{H}
  {\vert}D_{s}^{\ast}{\rangle}
   \nonumber \\ &=&
   \frac{G_{F}}{\sqrt{2}}\, V_{cs}^{\ast}\,V_{ud}\,
   a_{1}\, {\langle}{\pi}{\vert}(\bar{u}\,d)_{H}
  {\vert}0{\rangle}\, {\langle}{\phi}{\vert}
  (\bar{s}\,c)_{H} {\vert}D_{s}^{\ast}{\rangle}
   \label{Hamiltonian-amplitude},
   \end{eqnarray}
   where $(\bar{s}\,c)_{H}$ and $(\bar{u}\,d)_{H}$ are the
   color singlet $V$-$A$ hadron currents, and the subscript
   $H$ is introduced to indicate the change to hadron currents
   and distinguish with quark currents of
   Eq.(\ref{operator-o1}) and Eq.(\ref{operator-o2}).
   The effects from the color exchanges are embodied into the
   coefficient $a_{1}$ $=$ $C_{1}$ $+$ ${\xi}\,C_{2}$.
   It is expected ${\xi}$ $=$ $1/N_{c}$ $=$ $1/3$ from
   color matching.
   ${\xi}$ or $a_{1}$ sometimes is regarded as a parameter
   for different factorization approaches, because of the
   uncertain contributions of color octet current product
   and nonfactorizable contributions.
   The approximation of $a_{1}$ ${\approx}$ $1.1$ is frequently
   used in many phenomenological studies of nonleptonic decays
   for charmed hadron mesons, such as Refs.
   \cite{Stech.1985,npb.133.315,cpc.26.665,cpc.27.759,
   epjc.42.391,jpg.34.637,PhysRevD.81.074021,PhysRevD.84.074019,
   PhysRevD.86.014014,PhysRevD.86.036012,ijmpa.30.1550094,
   PhysRevD.100.093002}.

   Using the parameterization for amplitude in
   Eq.(\ref{Hamiltonian-amplitude}), the decay widths can
   be given in terms of measurable physical HMEs.
   The HMEs of hadron currents in Eq.(\ref{Hamiltonian-amplitude})
   are related to the decay constants and hadron transition form
   factors. The one-body HMEs are relevant to decay constants
   of hadrons,
   \begin{eqnarray}
  {\langle}0{\vert} \bar{d}\,{\gamma}_{\mu}\,u
  {\vert}{\pi}^{+}(p){\rangle} &=& 0
   \label{eq:hme-pseudoscalar-v},
   \\
  {\langle}0{\vert} \bar{d}\,{\gamma}_{\mu}\, {\gamma}_{5}\,u
  {\vert}{\pi}^{+}(p){\rangle} &=& i\,f_{\pi}\,p_{\mu}
   \label{eq:hme-pseudoscalar-a}.
   \end{eqnarray}
   The charged pion decay constant has been well determined
   from numerical lattice QCD simulations,
   $f_{\pi}$ $=$ $130.2{\pm}1.2$ MeV (See Ref. \cite{pdg2020}
   for a summary review).
   With the conventions of Refs. \cite{jhep.1912.102},
   the form factors are defined as,
    \begin{eqnarray} & &
   {\langle}{\phi}({\epsilon}_{2},p_{2}){\vert}\,
    \bar{s}\,{\gamma}_{\mu}\,c\,
   {\vert}D_{s}^{\ast}({\epsilon}_{1},p_{1}){\rangle}
    \nonumber \\ &=&
   -({\epsilon}_{1}{\cdot}{\epsilon}_{2}^{\ast})\,
    \big\{ P_{\mu}\,V_{1}(q^{2})
          -q_{\mu}\,V_{2}(q^{2}) \big\}
   -({\epsilon}_{1}{\cdot}q)\,
     {\epsilon}_{2,{\mu}}^{\ast}\,V_{5}(q^{2})
   +({\epsilon}_{2}^{\ast}{\cdot}q)\,
     {\epsilon}_{1,{\mu}}\,V_{6}(q^{2})
    \nonumber \\ & &
   +\frac{ ({\epsilon}_{1}{\cdot}q)\,
           ({\epsilon}_{2}^{\ast}{\cdot}q) }
         { m_{D_{s}^{\ast}}^{2}-m_{{\phi}}^{2} }\,
    \big\{ \big[ P_{\mu} -
    \frac{ m_{D_{s}^{\ast}}^{2}-m_{{\phi}}^{2} }{ q^{2} }\,q_{\mu}
    \big]\, V_{3}(q^{2})
   +\frac{ m_{D_{s}^{\ast}}^{2}-m_{{\phi}}^{2} }{ q^{2} }\,
           q_{\mu}\, V_{4}(q^{2}) \big\}
    \label{kine-v2v-v},
    \end{eqnarray}
    \begin{eqnarray} & &
   {\langle}{\phi}({\epsilon}_{2},p_{2}){\vert}\,
    \bar{s}\,{\gamma}_{\mu}\,{\gamma}_{5}\,c\,
   {\vert}D_{s}^{\ast}({\epsilon}_{1},p_{1}){\rangle}
    \nonumber \\ &=&
   -i\,{\varepsilon}_{{\mu}{\nu}{\alpha}{\beta}}\,
   {\epsilon}_{1}^{\alpha}\,
   {\epsilon}_{2}^{{\ast}{\beta}}\,\big\{ \big[ P^{\nu} -
    \frac{ m_{D_{s}^{\ast}}^{2}-m_{{\phi}}^{2} }{ q^{2} }\,q^{\nu}
    \big]\, A_{1}(q^{2})
   +\frac{ m_{D_{s}^{\ast}}^{2}-m_{{\phi}}^{2} }{ q^{2} }\,
    q^{\nu}\, A_{2}(q^{2}) \big\}
    \nonumber \\ & &
   -\frac{ i\,{\varepsilon}_{{\mu}{\nu}{\alpha}{\beta}}\,
           P^{\alpha}\, q^{\beta} }
         { m_{D_{s}^{\ast}}^{2}-m_{{\phi}}^{2} }\, \big\{
    ({\epsilon}_{2}^{\ast}{\cdot}q)\,
   {\epsilon}_{1}^{\nu}\,A_{3}(q^{2})
  - ({\epsilon}_{1}{\cdot}q)\,
   {\epsilon}_{2}^{{\ast},{\nu}}\,A_{4}(q^{2}) \big\}
    \label{kine-v2v-a},
    \end{eqnarray}
   where ${\epsilon}_{i}$ denotes the polarization vector
   of the vector mesons. The momentum
   $P$ $=$ $p_{1}$ $+$ $p_{2}$ and $q$ $=$ $p_{1}$ $-$ $p_{2}$.
   At the pole $q^{2}$ $=$ $0$, there is,
    \begin{equation}
    V_{3}(0)\, =\, V_{4}(0)
    \label{eq:v3ev4},
    \end{equation}
    \begin{equation}
    A_{1}(0)\, =\, A_{2}(0)
    \label{eq:a1ea2}.
    \end{equation}
   The values of formfactors for the $D_{s}^{\ast}$ ${\to}$
   ${\phi}$ transition have been obtained with the light
   front approach \cite{jhep.1912.102}, for example,
   $A_{1}(0)$ $=$ $0.65$, $V_{1}(0)$ $=$ $0.71$,
   $V_{4}(0)$ $=$ $0.28$, $V_{5}(0)$ $=$ $1.54$,
   and $V_{6}(0)$ $=$ $0.86$.

   Finally, the decay amplitude can be expressed by three
   invariant amplitudes.
   They are defined by the decomposition,
    \begin{eqnarray} & &
   {\cal A}(D_{s}^{\ast}{\to}{\phi}{\pi})
    \nonumber \\ &=&
     a\,({\epsilon}_{D_{s}^{\ast}}{\cdot}{\epsilon}_{\phi}^{\ast})
    +\frac{ b }{ m_{D_{s}^{\ast}}\,m_{\phi} }\,
        ({\epsilon}_{D_{s}^{\ast}}{\cdot}p_{\pi})\,
        ({\epsilon}_{\phi}^{\ast}{\cdot}p_{\pi})
    +\frac{ c }{ m_{D_{s}^{\ast}}\,m_{\phi} }\,
        {\varepsilon}_{{\mu}{\nu}{\alpha}{\beta}}\,
        {\epsilon}_{D_{s}^{\ast}}^{\alpha}\,
        {\epsilon}_{\phi}^{{\ast}{\beta}}\,
        p_{\pi}^{\mu}\, (p_{D_{s}^{\ast}}+p_{\phi})^{\nu}
    \nonumber \\ &=&
   {\epsilon}_{D_{s}^{\ast}}^{\alpha}\,
   {\epsilon}_{\phi}^{{\ast}{\beta}}\, \big\{
     a\,g_{{\alpha}{\beta}}
    +\frac{ b }{ m_{D_{s}^{\ast}}\,m_{\phi} }\,
       p_{{\pi},{\alpha}}\,
       p_{{\pi},{\beta}}\,
    +\frac{ c }{ m_{D_{s}^{\ast}}\,m_{\phi} }\,
        {\varepsilon}_{{\mu}{\nu}{\alpha}{\beta}}\,
        p_{\pi}^{\mu}\, (p_{D_{s}^{\ast}}+p_{\phi})^{\nu}  \big\}
    \label{decay-amplitude},
    \end{eqnarray}
   and the invariant amplitudes $a$, $b$, and $c$ describe
   the $s$-, $d$-, and $p$-wave contributions.
    \begin{eqnarray}
    a &=&
      -i\,\frac{G_{F}}{\sqrt{2}}\, V_{cs}^{\ast}\,V_{ud}\,
      f_{\pi}\, (m_{D_{s}^{\ast}}^{2}-m_{{\phi}}^{2})\,
      a_{1}\, V_{1}(0)
    \label{decay-amplitude-a}, \\
    b &=&
      -i\,\frac{G_{F}}{\sqrt{2}}\, V_{cs}^{\ast}\,V_{ud}\,
      f_{\pi}\,m_{D_{s}^{\ast}}\,m_{\phi}\,a_{1}\,
      \big\{ V_{5}(0)-V_{6}(0)-V_{4}(0) \big\}
    \label{decay-amplitude-b}, \\
    c &=&
      - \frac{G_{F}}{\sqrt{2}}\, V_{cs}^{\ast}\,V_{ud}\,
      f_{\pi}\,m_{D_{s}^{\ast}}\,m_{\phi}\,a_{1}\, A_{1}(0)
    \label{decay-amplitude-c}.
    \end{eqnarray}

   In the rest frame of the $D_{s}^{\ast}$ meson,
   branching ratio is defined as,
    \begin{eqnarray}
   {\cal B}(D_{s}^{\ast}{\to}{\phi}{\pi})
   &=&
    \frac{1}{24\,{\pi}}\,
    \frac{p_{\rm c.m.}}{m_{D_{s}^{\ast}}^{2}\,{\Gamma}_{D_{s}^{\ast}}}
   {\vert}{\cal A}(D_{s}^{\ast}{\to}{\phi}{\pi}){\vert}^{2}
   \nonumber \\ &=&
    \frac{1}{24\,{\pi}}\,
    \frac{p_{\rm c.m.}}{m_{D_{s}^{\ast}}^{2}\,{\Gamma}_{D_{s}^{\ast}}}
    \big\{ {\vert}a{\vert}^{2}\,(2+x^{2})
         + {\vert}b{\vert}^{2}\,(x^{2}-1)^{2}
    \nonumber \\ & &
    + {\vert}2\,c{\vert}^{2}\,2\,(x^{2}-1)
    - 2\,{\rm R}e(a\,b^{\ast})\,x\,(x^{2}-1) \big\}
    \label{eq:branch},
    \end{eqnarray}
   where the center-of-mass momentum of final states is of magnitude,
    \begin{equation}
    p_{\rm c.m.}\, =\, \displaystyle
    \frac{ {\lambda}^{\frac{1}{2}}(
           m_{D_{s}^{\ast}}^{2},
           m_{\phi}^{2},
           m_{\pi}^{2}) }
         { 2\,m_{D_{s}^{\ast}} }
    \label{pcm},
    \end{equation}
   the parameter $x$ is defined as,
    \begin{equation}
    x\, =\, \frac{ p_{D_{s}^{\ast}}{\cdot}p_{\phi} }
                 { m_{D_{s}^{\ast}}\,m_{\phi} }
     \, =\, \frac{ E_{\phi} }{ m_{\phi} }
     \, =\, \frac{ m_{D_{s}^{\ast}}^{2}+m_{\phi}^{2}-m_{\pi}^{2}  }
                 { 2\,m_{D_{s}^{\ast}}\,m_{\phi} }
    \label{x},
    \end{equation}
    \begin{equation}
    {\lambda}(x,y,z)\, =\, x^{2}+y^{2}+z^{2}-2\,x\,y-2\,y\,z-2\,z\,x
    \label{lambda},
    \end{equation}
    \begin{equation}
    p_{\rm c.m.}^{2}\, =\, m_{\phi}^{2}\,(x^{2}-1)
    \label{pcm-x}.
    \end{equation}

   \section{numerical results and discussion}
   \label{sec:result}

   The total decay width ${\Gamma}_{D_{s}^{\ast}}$ $<$
   $1.9$ MeV was set at the 90\% confidence level by the
   CLEO collaboration in 1995 \cite{PhysRevLett.75.3232}.
   A quantitative and concrete result currently comes from
   theoretical estimations. Because of the lion's share
   ${\cal B}(D_{s}^{\ast}{\to}{\gamma}D_{s})$
   $=$ $(93.5{\pm}0.7)\%$ \cite{pdg2020}, an approximation
   ${\Gamma}_{D_{s}^{\ast}}$ ${\approx}$
   ${\Gamma}(D_{s}^{\ast}{\to}{\gamma}D_{s})$
   is often used in theoretical calculation.
   The decay width for the magnetic dipole transition
   is \cite{epjc.81.1110},
   \begin{equation}
  {\Gamma}(D_{s}^{\ast}{\to}D_{s}{\gamma})\, =\,
   \frac{4}{3}\,{\alpha}_{\rm em}\,k_{\gamma}^{3}\,
  {\mu}_{D_{s}^{\ast}D_{s}}^{2}\,\, {\approx}\,
   0.36\,\text{keV}
   \label{eq:decay-width-ds},
   \end{equation}
  where ${\mu}_{D_{s}^{\ast}D_{s}}$ is the magnetic dipole
  moment and $k_{\gamma}$ is the momentum of photon.

  Using Eq.(\ref{eq:branch}), we can obtain branching ratio,
   \begin{equation}
  {\cal B}(D_{s}^{\ast}{\to}{\phi}{\pi})\,
  {\approx}\, 2.4 {\times}\,
   \frac{0.36\,\text{keV}}{ {\Gamma}_{D_{s}^{\ast}} }\,
  {\times}\, 10^{-7}
   \label{eq:br-ds},
   \end{equation}
   and the corresponding partial decay width,
   ${\Gamma}( D_{s}^{\ast} {\to} {\phi}{\pi})$ ${\approx}$
   $0.86\,{\times}\, 10^{-13}$ GeV,
   is more than twice as large as the recent estimate using
   the QCD light cone sum rules in Ref. \cite{cheng2203}
   where a relatively smaller coefficient $a_{1}$
   ${\approx}$\, $1.0$ is used.

   We will make two comments on branching ratio.
   (1)
   There are many factors which influence the numerical results,
   such as the final state interactions.
   It is foreseeable that there will very large theoretical
   uncertainties.
   For example, using a much smaller decay width
   ${\Gamma}_{D_{s}^{\ast}}$ ${\approx}$ $0.07$ keV from the
   lattice QCD simulations \cite{PhysRevLett.112.212002},
   branching ratio will be increased five times.
   Our focus is whether there is feasible to explore the
   $D_{s}^{\ast}$ meson via the ${\phi}{\pi}$ final states at
   the future experiments.
   A rough estimate rather than precise calculation on
   branching ratio is enough.
   (2)
   For the tree-dominated and color-favored nonleptonic
   heavy flavored meson decays arising from the external
   $W$ emission weak interaction, there is a consensus
   that NF approximation does hold and can give a reasonable
   and correct magnitude order estimation on branching ratio.
   In this sence, ${\cal B}(D_{s}^{\ast}{\to}{\phi}{\pi})$
   ${\sim}$ ${\cal O}(10^{-7})$ seems credible.

   Based on the above analysis, it can be conclude that
   the $D_{s}^{\ast}$ ${\to}$ ${\phi}{\pi}$ decay
   should be measurable in the future experiments, such
   as STCF, SuperKEKB, CEPC, FCC-ee and LHCb.
   The potential event numbers of the $D_{s}^{\ast}$
   mesons and the $D_{s}^{\ast}$ ${\to}$ ${\phi}{\pi}$
   decays are listed in Table \ref{tab:number-ds}.
   It is clearly seen from Table \ref{tab:number-ds} that
   the natural properties of the $D_{s}^{\ast}$ meson can
   be investigated via the $D_{s}^{\ast}$ ${\to}$
   ${\phi}{\pi}$ weak decays, particularly in the future
   FCC-ee and LHCb experiments.

  \begin{table}[th]
  \caption{The potential event numbers of the $D_{s}^{\ast}$
  meson available and the $D_{s}^{\ast}$ ${\to}$ ${\phi}{\pi}$
  decays in the future experiments,
  with the branching ratio ${\cal B}(Z^{0}{\to}c\bar{c})$
  ${\approx}$ $12\%$ \cite{pdg2020} and
  ${\cal B}(D_{s}^{\ast}{\to}{\phi}{\pi})$ ${\approx}$
  $3\,{\times}\, 10^{-7}$, the fragmentation
  fraction $f(c{\to}D_{s}^{\ast})$ ${\approx}$ $5.5\%$
  \cite{epjc.76.397} and the identification
  efficiency ${\epsilon}$ ${\sim}$ $20\%$.}
  \label{tab:number-ds}
  \begin{ruledtabular}
  \begin{tabular}{c|cccl}
     experiment
   & $N_{D_{s}^{\ast}}$
   & $N_{D_{s}^{\ast}{\to}{\phi}{\pi}}$
   & ${\epsilon}{\times}N_{D_{s}^{\ast}{\to}{\phi}{\pi}}$
   & \multicolumn{1}{c}{remarks} \\ \hline
     STCF \cite{STCF,SCTF}
   & $10^{10}$ \cite{epjc.81.1110}
   & $3000$ & $600$
   & with $10\,ab^{-1}$ data  \\
     SuperKEKB \cite{PTEP.2019.123C01}
   & $5.5{\times}10^{9}$
   & $1600$ & $300$
   & with $5{\times}10^{10}$ charm quark pairs  \\
     CEPC \cite{cepc}
   & $1.3{\times}10^{10}$
   & $4000$ & $800$
   & from $10^{12}$ $Z^{0}$ boson decays \\
     FCC-ee \cite{fcc}
   & $6.6{\times}10^{10}$
   & $2{\times}10^{4}$
   & $4000$
   & from $5{\times}10^{12}$ $Z^{0}$ boson decays \\
     LHCb@HL-LHC \cite{epjst.228.1109}
   & $4{\times}10^{13}$
   & $10^{7}$  & $2{\times}10^{6}$
   & with $300\,fb^{-1}$ data
   \end{tabular}
   \end{ruledtabular}
   \end{table}

  \section{Summary}
  \label{sec:summary}
  Inspired by the inadequate understanding of the properties
  of $D_{s}^{\ast}$ meson, and the promisingly experimental
  prospects of investigating the $D_{s}^{\ast}$ meson in the
  future high-luminosity experiments, the $D_{s}^{\ast}$
  ${\to}$ ${\phi}{\pi}$ decay was studied by using
  the NF approach within SM.
  The nonleptonic $D_{s}^{\ast}$ ${\to}$ ${\phi}{\pi}$ weak decay offers
  a fresh arena and a tempting opportunity to explore the
  wanted $D_{s}^{\ast}$ meson, although with a very tiny
  occurrence probability of ${\sim}$ ${\cal O}(10^{-7})$.
  The final states of the $D_{s}^{\ast}$ ${\to}$ ${\phi}{\pi}$
  decay have the relatively larger momenta than those
  of the predominant electromagnetic decays $D_{s}^{\ast}$
  ${\to}$ $D_{s}{\gamma}$ and ${\to}$ $D_{s}{\pi}$,
  and can be more easily identified by the sensitive
  detectors.
  It is found that several thousands of events for
  the $D_{s}^{\ast}$ ${\to}$ ${\phi}{\pi}$ decay are
  expected to be accessible at the STCF,
  SuperKEKB, CEPC and FCC-ee experiments,
  several millions of events at LHCb@HL-LHC experiments.
  It is practicable to experimentally study the
  $D_{s}^{\ast}$ ${\to}$ ${\phi}{\pi}$ weak decay
  in the future.

  \section*{Acknowledgments}
  The work is supported by the National Natural Science Foundation
  of China (Grant Nos. 11705047, U1632109, 11875122) and Natural
  Science Foundation of Henan Province (Grant No. 222300420479),
  the Excellent Youth Foundation of Henan Province
  (Grant No. 212300410010).

  \end{document}